\documentclass[conference, letterpaper, fleqn]{IEEEtran}
\ifCLASSINFOpdf
\else
\fi
\ifCLASSINFOpdf
   \usepackage[pdftex]{graphicx}
\else
\fi

\usepackage[cmex10]{amsmath}
\usepackage{color}
\usepackage{fancyhdr}
\usepackage[caption=false,font=footnotesize]{subfig}

\renewcommand{\thispagestyle}[2]{}

\fancypagestyle{plain}{
        \fancyhead{}
        \fancyhead[C]{first page center header}
        \fancyfoot{}
        \fancyfoot[C]{first page center footer}
}
\usepackage[pdftex]{graphicx}

\newtheorem{theorem}{Theorem}
\newtheorem{lemma}{Lemma}

\newtheorem{definition}{Definition}

\newcommand{\QEDPR}{\hspace*{\fill}\rule{0.6em}{0.6em}}

\newcommand{\ST}{\mbox{{\tt ST}}}
\newcommand{\A}{\mbox{{$\cal A$}}}
\newcommand{\FOR}{\mbox{{\bf for}}}
\newcommand{\TO}{\mbox{{\bf to}}}
\newcommand{\DO}{\mbox{{\bf do}}}

\begin{document}

%
\title{Solving Dynamic Programming Problem by\\
  Pipeline Implementation on GPU}

   \author{
     \IEEEauthorblockN{Susumu Matsumae}
     \IEEEauthorblockA{Department of Information Science\\
       Graduate School of Science and Engineering\\
       Saga University\\
       Saga, Japan}
     \and
     \IEEEauthorblockN{Makoto Miyazaki}
     \IEEEauthorblockA{Department of Information Science\\
       Graduate School of Science and Engineering\\
       Saga University\\
       Saga, Japan}
   }

\maketitle

\begin{abstract}
  In this paper, we show the effectiveness of a pipeline
  implementation of Dynamic Programming (DP) on GPU.  As an example,
  we explain how to solve a matrix-chain multiplication (MCM) problem
  by DP on GPU.  This problem can be sequentially solved in $O(n^3)$
  steps by DP where $n$ is the number of matrices, because its
  solution table is of size $n\times n$ and each element of the table
  can be computed in $O(n)$ steps.  A typical speedup strategy for
  this is to parallelize the $O(n)$ step computation of each element,
  which can be easily achieved by parallel prefix computation, i.e.,
  an $O(\log n)$ step computation with $n$ threads in a tournament
  fashion.  By such a standard parallelizing method, we can solve the
  MCM problem in $O(n^2 \log n)$ steps with $n$ threads.  In our
  approach, we solve the MCM problem on GPU in a pipeline fashion,
  i.e., we use GPU cores for supporting pipeline-stages so that many
  elements of the solution table are partially computed in parallel at
  one time. Our implementation determines one output value per one
  computational step with $n$ threads in a pipeline fashion and
  constructs the solution table totally in $O(n^2)$ steps with $n$
  threads.
\end{abstract}


%
\IEEEpeerreviewmaketitle

\section{Introduction}

In this paper, we show the effectiveness of a pipeline implementation
of Dynamic Programming (DP) on GPU.  As an example, we explain how to
solve a {\it matrix-chain multiplication (MCM)} problem
\cite{Cormen:2009:IAT:1614191} by DP on GPU.  This problem can be
sequentially solved in $O(n^3)$ steps by DP where $n$ is the number of
matrices, because its solution table is of size $n\times n$ and each
element of the table can be computed in $O(n)$ steps.  A typical
speedup strategy for this is to parallelize the $O(n)$ step
computation of each element, which can be easily achieved by parallel
prefix computation, i.e., an $O(\log n)$ step computation with $n$
threads in a tournament fashion.  By such a standard parallelizing
method, we can solve the MCM problem in $O(n^2 \log n)$ steps with $n$
threads.

It has been studied well to speed up DP programs using GPU
(e.g. \cite{YasuakiITO2013,nakano-time-opt-hmm}), where they mainly
focus on optimizing the order of accessing data by proposing novel
techniques avoiding memory access conflicts.  In this study, we
consider adopting a pipeline technique and implementing the DP program
on GPU in a pipeline fashion.  The pipeline computation technique
\cite{roosta1999} can be used in situations in which we perform
several operations $\{OP_1,OP_2,\ldots,OP_n\}$ in a sequence, where
some steps of each $OP_{i+1}$ can be carried out before operation
$OP_i$ is finished.  In parallel algorithms, it is often possible to
overlap those steps and improve total execution time.

In our approach, we solve the MCM problem on GPU in a pipeline
fashion, i.e., we use GPU cores for supporting pipeline-stages so that
many elements of the solution table are partially computed in parallel
at one time. Our implementation determines one output value per one
computational step with $n$ threads in a pipeline fashion and
constructs the solution table totally in $O(n^2)$ steps with $n$
threads.  This paper is an extended version of our conference paper
\cite{miyazaki2018}.

The rest of this paper is organized as follows.  Section II introduces
problem definitions and base algorithms.  Section III explains our
pipeline implementations for DP on GPU and offers some experimental
results.  Section IV explains how to apply the pipeline implementation
technique to the MCM problem, and finally Section V offers concluding
remarks.

\section{Preliminaries}
In this section, we introduce some preliminary definitions and base
algorithms.  We first define a simplified DP problem to be solved on
GPU, and then explain our GPU implementations of programs.

\subsection{Simplified DP Problem}

In this study, we implement a typical DP program on GPU.  To simplify
the exposition, we focus on programs that solve such a simplified DP
problem defined as follows:
\begin{definition}\label{def:1}{\it (Simplified DP Problem)}\ \ 
  A one-dimensional array \ST$[0, \ldots, n-1]$ of size $n$ as a
  solution table, a set $\A=\{a_1,a_2,\ldots,a_{k}\}$ of $k$ integers
  representing offset numbers, and a semi-group binary operator
  $\otimes$ over integers are given.  Every element of set \A\
  satisfies the following inequality:
  \[ a_1>a_2>\cdots>a_k>0.\]
  Then, a {\it simplified DP problem} (S-DP problem) is to fill all
  the elements of array \ST\ in such a way that each \ST$[i]$ is
  computed by the following equation:
  \begin{equation}
    \label{eq:0}
    \ST[i] = \otimes_{1\leq j \leq k}\; \ST[i-a_j]
  \end{equation}
  where
  \ST$[0]$, \ST[1],$\ldots$ ,\ST$[a_1-1]$ are preset with initial
  values. \QEDPR
\end{definition}

For example, Fibonacci number problem is equal to the S-DP problem
where $k=2, a_1=2, a_2=1$, $\otimes=+$, and \ST$[0]$=\ST$[1]$=$1$.

\subsection{Conventional Approach to S-DP Problem}

To begin with, we show a straightforward sequential algorithm that
solves the S-DP problem.  Fig.\ \ref{fig:DPalgo} shows the algorithm.
\begin{figure}[h]
  {\hrulefill
    \ \\

    \ \ \underline{\bf A Sequential Algorithm for S-DP Problem}\\

    \ \ \ \FOR\ {\tt i} = $a_1$ \TO\ $n-1$ \DO
    \begin{quote}
      \ST{\tt [i]} = \ST{\tt[i}$-a_1${\tt ]}; \\
      \FOR\ {\tt j} = $2$ \TO\ $k$ \DO
      \begin{quote}
        \ST{\tt [i]} = \ST{\tt [i]}$\;\otimes\;$\ST{\tt [i}$-a_{\tt
          j}${\tt ]};
      \end{quote}
    \end{quote}
  \hrulefill }
\caption{A sequential algorithm for S-DP problem}
  \label{fig:DPalgo}
\end{figure}\\
The outer loop computes values from \ST$[a_1]$ to \ST$[n-1]$ in order,
and the inner loop computes \ST{\tt [i]} for each {\tt i} by equation
(\ref{eq:0}).  Since the outer loop takes $n-a_1+1 = O(n)$ steps and the
inner loop requires $O(k-1)$ steps, this algorithm takes $O(nk)$ steps
in total.

Next, we consider parallelizing the algorithm for S-DP problem.  The
straightforward approach is to parallelize the inner loop by using GPU
cores.  We can easily write a multi-thread program that executes the
inner loop-body, $\ST{\tt [i]}=\ST{\tt [i]}\;\otimes\;\ST{\tt
  [i}-a_{\tt j}{\tt ]}$, for each j
in parallel using $k-1$ threads at one time.  Such an implementation,
however, does not improve the time cost, because every thread has
access to the same {\tt \ST[i]} and thus memory access conflicts occur.  As
a result, those memory conflicts should be automatically solved at
run-time by the serializing mechanism of GPU, and consequently the
whole time cost stays in $O(nk)$ steps, which is the same time cost as
that of the sequential implementation.

To avoid the memory access conflicts, we can use a well-known standard
parallel prefix computation algorithm
\cite{Leighton:1991:IPA:119339,Kumar:1994:IPC:156619}, in which the
computations of $\;\otimes\;$ over $k$ values are executed in a tournament
fashion.  Since the parallel prefix computation runs in $O(\log
k)$ steps for $k$ values, obviously the entire time cost can be
improved to $O(n\log k)$ steps by using $k$ threads.

Although we can successfully reduce the time cost from $O(nk)$ to $O(n
\log k)$ by using the parallel prefix computation, it is not
work-time optimal because there are many idle threads during the
computations in a tournament fashion.  In the next section we propose
other parallel implementation strategy and show that we can improve
the time cost further.

\section{Pipeline Implementation on GPU}
In this section, we explain our proposed parallel implementations for
S-DP problem on GPU.  Our program runs in a pipeline fashion.

\subsection{Pipeline Implementation for S-DP problem}
In our implementation, we use a group of $k$ threads to establish
$k$-stage pipeline, and this thread group treats $k$ consecutive
elements at one time in parallel.  Fig.\ \ref{fig:pipeline-alg}
describes our pipeline algorithm for the S-DP problem.  The index
variable {\tt i} of the outer loop stands for the head position of the
$k$-thread group.  The inner loop controls each thread's behaviour in
such a way that the {\tt j}-th thread executes computation for {\tt
  \ST[i-j+1]} using the value stored in {\tt \ST[i-j+1-$a_{\tt j}$]}.

\begin{figure}[h]
  {\hrulefill
    \ \\

    \underline{\bf A Pipeline Algorithm for S-DP Problem}\\
  
    \ \ \ \FOR\ {\tt i} = $a_1$ \TO\ $n+k-2$ \DO
    \begin{quote}
      \FOR\ {\tt j} = $1$ \TO\ $k$ \DO\ \underline{\bf in parallel}
      \begin{quote}
        Thread {\tt j} executes the following operation if $a_1\leq i_{\tt j} <
        n$ where $i_{\tt j}={\tt i}-{\tt j}+1$:
         \[
         \ST[i_{\tt j}] = \begin{cases}
           \ST[i_{\tt j}-a_{\tt j}]; \;\;({\tt j}=1) \\
           \ST[i_{\tt j}] \;\otimes\; \ST[i_{\tt j}-a_{\tt j}]; \;\; ({\tt j}>1)
         \end{cases}
         \]
      \end{quote}
    \end{quote}
  \hrulefill
  }
  \caption{A pipeline algorithm for S-DP problem}
  \label{fig:pipeline-alg}
\end{figure}

\begin{figure*}[t]
    \ \\

  \centering
    \ \\

  \includegraphics[width=16cm]{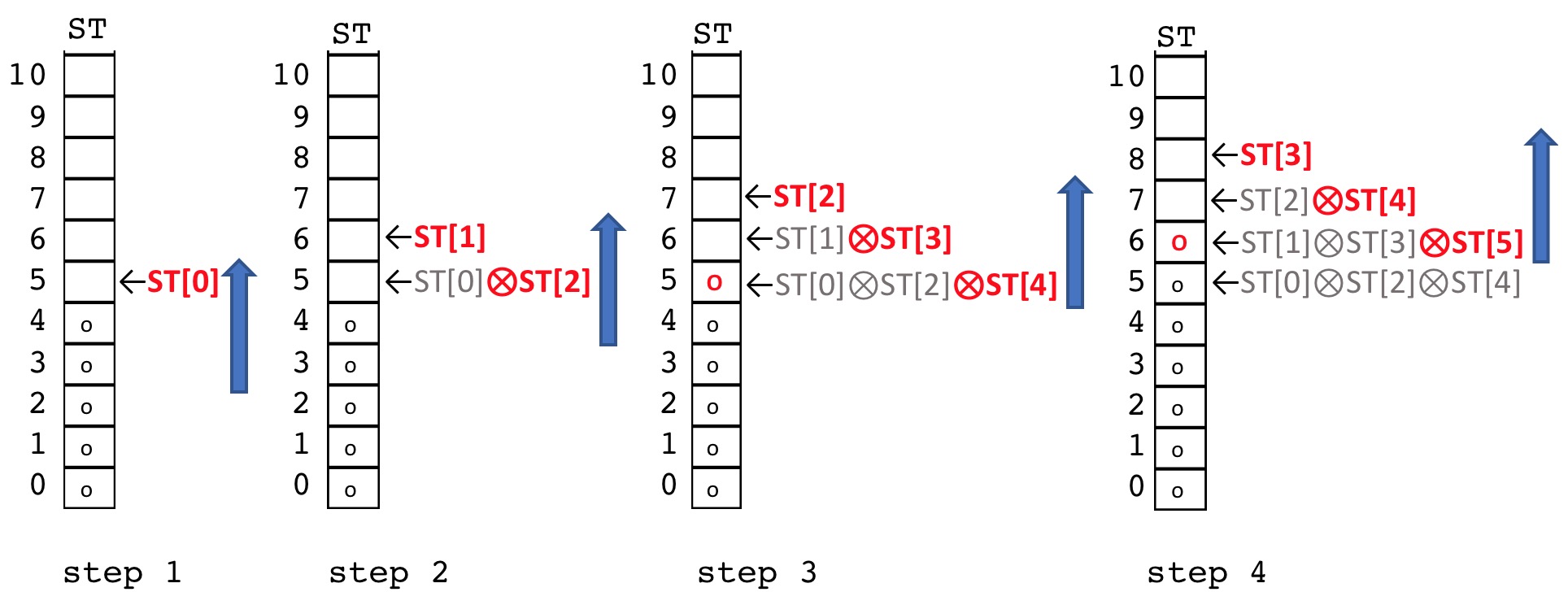}
  \caption{An execution example for the case where $k=3$, $a_1=5$, $a_2=3$, and
    $a_3=1$ hold and initial values are preset to \ST$[0]$, \ST[1],
    $\ldots$, and \ST$[4]$.}
  \label{fig:fig-pipe1}

    \ \\

\end{figure*}

An execution example is shown in Fig.\ \ref{fig:fig-pipe1}, where
$k=3$, $a_1=5$, $a_2=3$, and $a_3=1$ hold and the initial values are
already stored in \ST[0], \ST[1],$\ldots$ , \ST[4].
In Step 1, the head position {\tt i} of the thread group is 5.  In this step
the only one thread is activated and executes \ST[5] $\leftarrow$
\ST[0].  In Step 2, the head position is incremented to 6, and two
threads are activated.  The first thread treats \ST[6] and the second
thread works on \ST[5].  In Step 3, the head position becomes 7, and
now all $k=3$ threads actively execute operations for \ST[7], \ST[6],
and \ST[5] respectively.  It should be noted that finally in Step 3
the content of \ST[5] is completely determined while those of \ST[7]
and \ST[6] are partially computed and not yet determined.  From Step
3, all the $k=3$ threads are active until Step $n-a_1$ when the head
position {\tt i} of thread group reaches $n-1$, and after that step the
number of active threads decreases one by each step.  As you can see
there is no memory access conflict in this example.

As for the time-complexity of our pipeline implementation, from a
theoretical viewpoint, it takes only $O(n)$ steps, because the outer
loop takes $n+k-a_1-1=O(n)$ cycles and the inner loop requires $O(1)$ time
if there is no adjacent offset pair $(a_m, a_{m+1})$ such that
$a_m=a_{m+1}+1$.

\begin{figure}[h]
    \ \\

  \centering
  \includegraphics[width=7cm]{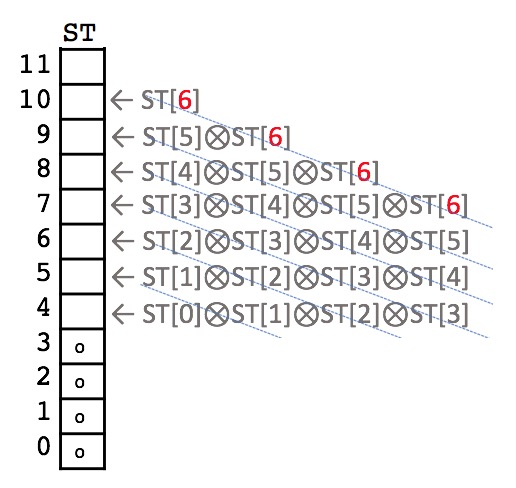}
   \caption{An example for the worst case where the offset numbers are
     consecutively given.  In this example, we have $k=4$, $a_1=4$,
     $a_2=3$, $a_3=2$, and $a_4=1$.}
   \label{fig:fig-pipe2}

    \ \\

 \end{figure}

However, from a practical viewpoint, because of the memory access
conflicts, the inner loop may take more time steps.  Actually, in the
worst case when consecutive offset numbers are given, those
\ST$[i_{\tt j}-a_{\tt j}]$, in the right hand side of the assignment
statement, coincidentally become the same element of array \ST\, and
hence the worst memory access conflicts occur.  In such a case, all
threads in the inner loop are serialized and it takes time
proportional to $k$.  See Fig.\ \ref{fig:fig-pipe2} for such a worst
case example.  In this example, all four threads try to have access to
\ST[${\tt i}-4$] at the same time in the inner loop.

Let $seq=(a_{p}, a_{p+1},\ldots,a_{q})$ be one of the longest
sub-sequences of given offset numbers $(a_1,a_2,\ldots,a_k)$
satisfying $a_r=a_{r+1}+1$ for all $p\leq r < q$.  Then, it is easy to
check that in the inner loop every thread $r$ ($p\leq r < q$) has
access to the same element of array \ST, and as a result those
conflicts are serialized at run time by GPU's serializing mechanism
and hence the memory access time becomes $(q-p+1)$ times slower than
that of conflict-free case.  For such a case, in \cite{miyazaki2018},
we proposed a {\it 2-by-2 pipeline implementation} technique where
each thread invoked in the inner loop executes two computations for
each element of array \ST.  The details can be found in
\cite{miyazaki2018}.

\subsection{Experimental Results}

Before going to the next section, we show the performance of our
pipeline implementation on GPU.  We use a computer with 3.40 [GHz]
Intel Xeon CPU E3-1245 v3 and an NVIDIA GeForce GTX TITAN Black.  The
OS is Ubuntu 17.10 and we use g++ 7.2.0 for compiling cpp programs and
CUDA 9.2 \cite{cuda9.2}.  The experimental results are shown in TABLE
\ref{tab:experiment}.

TABLE \ref{tab:experiment} shows the average execution time for each
implementation on GPU.  In the table, SEQUENTIAL, NAIVE-PARALLEL, and
PIPELINE respectively stand for the sequential implementation, the
naive multi-thread implementation, and our pipeline implementation
(for a general case).  Here, we use $\min$ operation for $\otimes$.
The average is computed among 100 executions for each setting.

\begin{table}[h]

  \caption{Execution time of Sequential, Naive parallel, and Pipeline
    implementations (msec)}
  \label{tab:experiment}
  \centering
  \begin{tabular}{|c||r|r|r|}
    \hline
    & SEQUENTIAL & NAIVE-PARALLEL & PIPELINE \\
    \hline
    \hline
    $2^{14} \leq n \leq 2^{15}$, &  & & \\
    $2^{12}\leq k \leq 2^{13}$ & 274 & 64 & 78\\
    \hline
    $2^{16} \leq n \leq 2^{17}$, &  &  & \\
    $2^{14}\leq k \leq 2^{15}$ & 4,288 & 368 & 386\\
    \hline
    $2^{18} \leq n \leq 2^{19}$, & & & \\
    $2^{16}\leq k \leq 2^{17}$ & 68,453 & 3,018 & 2,408\\
    \hline
  \end{tabular}

  \ \\

\end{table}

As for the comparison between the sequential implementation and the
parallel ones, parallel implementations are much faster even though it
is NAIVE-PARALLEL.  Although there is no difference in time between
NAIVE-PARALLEL and PIPELINE until $n \leq 2^{17}$, PIPELINE is faster
than NAIVE-PARALLEL when $n \geq 2^{18}$.

\section{Solving MCM Problem}

In this section, we explain how to apply our pipeline implementation
technique to more general DP problems.  As an example, we deal with
the matrix chain multiplication problem (MCM problem)
\cite{Cormen:2009:IAT:1614191}.

\subsection{Outline of Pipeline Implementation for MCM Problem}
\begin{figure}[h]
  \centering
  \includegraphics[width=6.5cm]{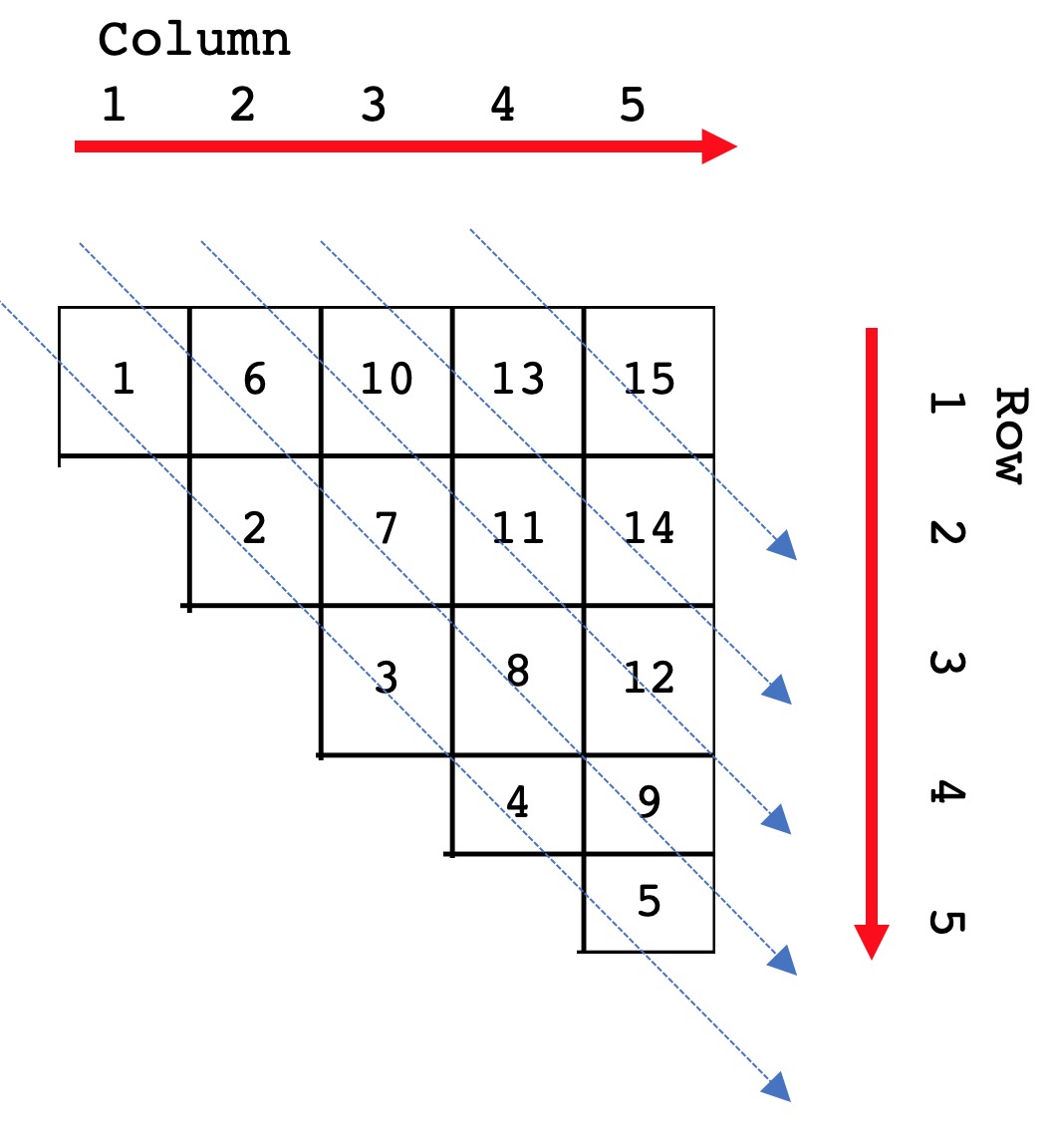}
  \caption{A solution table example for a matrix chain multiplication
    problem.}
  \label{fig:matrix-chain-st}
\end{figure}

It is well-known that the MCM problem can be efficiently solved by DP
with a two-dimensional solution table of triangular shape, and that
each element is computed along the diagonal direction.  See Fig.\
\ref{fig:matrix-chain-st} for an example.  In the figure, each number
represents the order of elements to be computed.

\begin{figure}[h]
  \centering
  \includegraphics[width=8cm]{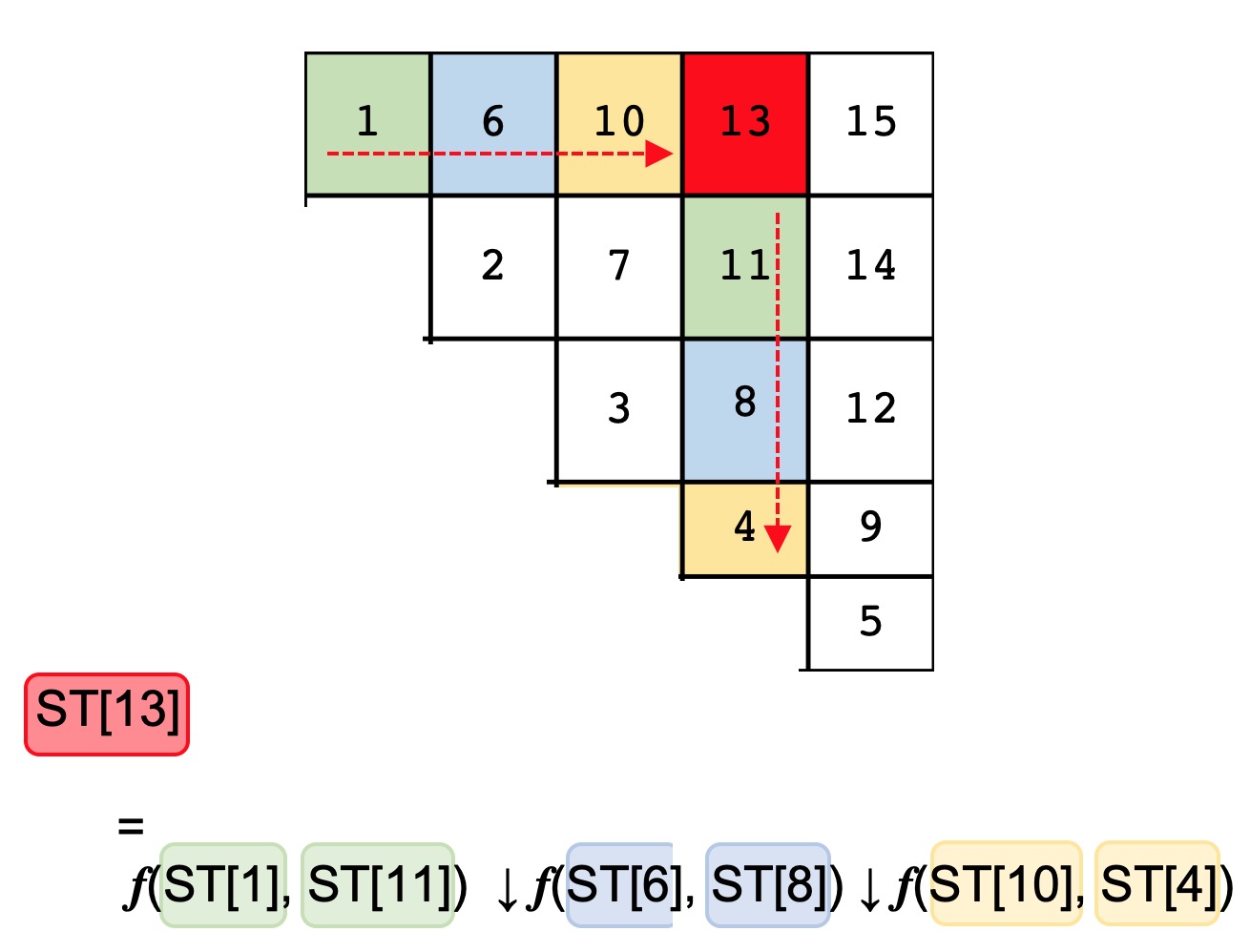}
  \caption{An example of computing an element of solution table.
    Here, we write \ST[$x$] for the element marked $x$.}
  \label{fig:matrix-chain-st2}
\end{figure}

The content of each element is computed by a binary operator
$\downarrow$ returning a smaller operand and a binary function
$f(*,*)$ as in Fig.\ \ref{fig:matrix-chain-st2}.  The detailed
definition of the MCM problem can be found in
\cite{Cormen:2009:IAT:1614191}.  In the figure, the element marked 13
is computed by the elements marked by 1, 6, 10, 11, 8, and 4.  If we
write \ST[$x$] for the element marked $x$ here, the computation is
expressed as\\
\[\ST[13]\]
\[=f(\ST[1],\ST[11])\downarrow
f(\ST[6],\ST[8])\downarrow
f(\ST[10],\ST[4]).\]

\begin{figure*}[t]

  \ \\
  \ \\
  \centering
  \includegraphics[width=12cm]{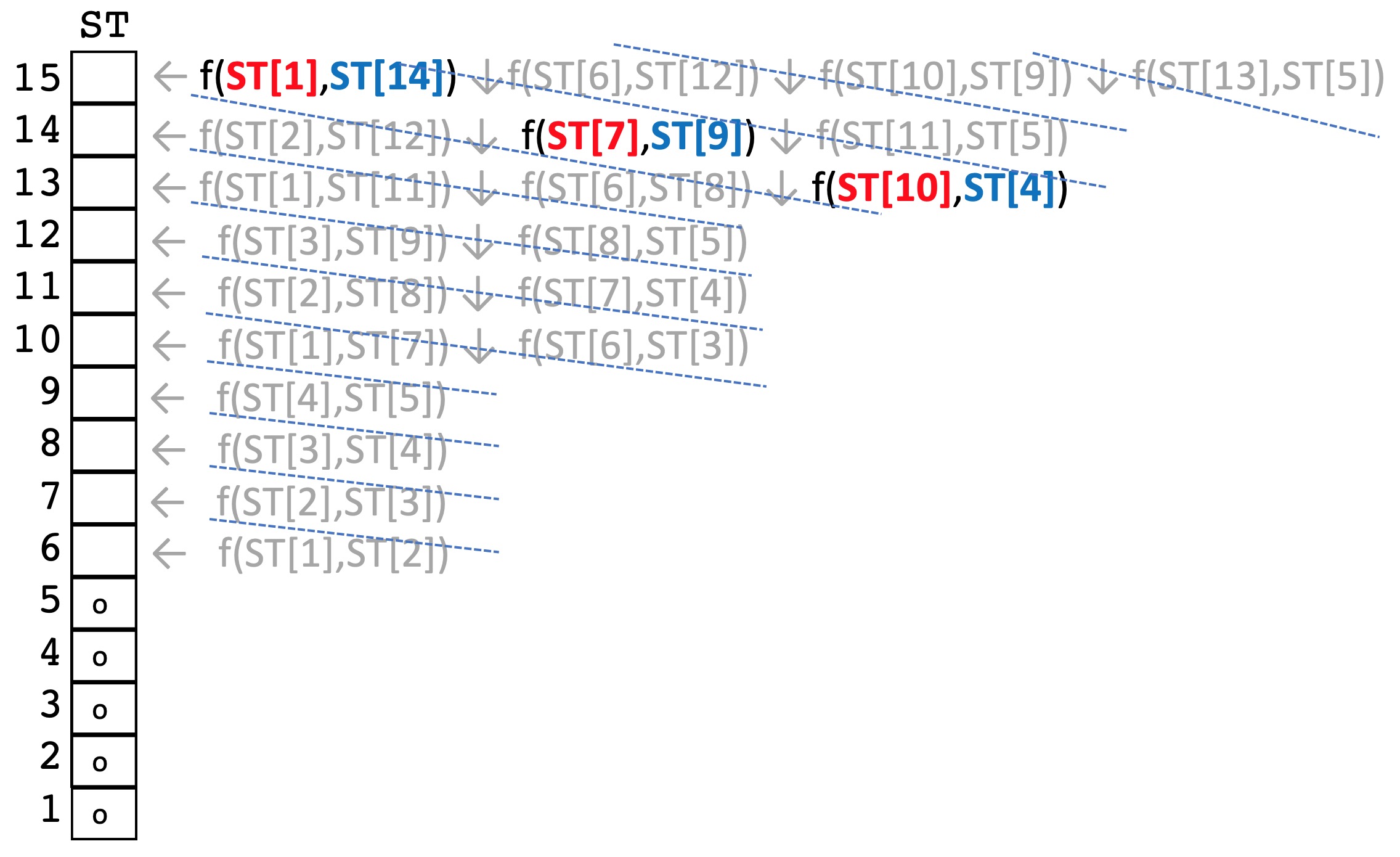}

   \caption{An execution example of pipeline implementation for the
     matrix chain multiplication problem.}
   \label{fig:matrix-chain-st3}

 \end{figure*}

Since the elements of two-dimensional solution table are computed in a
total order (linear order), we can line up them into a linear array
according to that order.  Once the solution table is transformed into
a linear array, we can apply the pipeline technique to the MCM problem
as well.  An execution example is shown in
Fig.\ \ref{fig:matrix-chain-st3}.

\subsection{Detailed Pipeline Implementation for MCM Problem}

Here, we explain how to solve the MCM problem by pipeline
implementation in details.

Firstly, we assume that the original two-dimensional solution table of
diagonal shape is already mapped to a one dimensional solution table
\ST\ appropriately.  That is, we line up all the elements of the
two-dimensional table into a linear array \ST\ according to the total
(linear) order in which each element is computed along the diagonal
direction as in Fig.\ \ref{fig:matrix-chain-st} and
\ref{fig:matrix-chain-st3}.  Since the number of input matrices is
$n$, the number of elements in the solution table is $n(n+1)/2$.
First $n$ elements of array \ST\ are preset with initial values, those
correspond to the elements located in the diagonal line of the
original two-dimensional table.

Let each computation for \ST[$i$] be represented as
\begin{equation}
  \label{eq:1}
  \ST[i] = {\downarrow\;}_{1\leq j \leq k_i}\;\;
  f(l_{(i,j)},r_{(i,j)}).
\end{equation}
It should be noted that here each $k_i$ may differ from others and
that it is $n-1$ at largest.
See Fig.\ \ref{fig:matrix-chain-st} and \ref{fig:matrix-chain-st3} for
an example.  Here, we have $n=5$, and \ST[13] and \ST[12] can be
represented as follows:
\[\ST[13]\]
\[=f(l_{(13,1)},r_{(13,1)}) \downarrow f(l_{(13,2)},r_{(13,2)}) \downarrow f(l_{(13,3)},r_{(13,3)})\]
\[=f(\ST[1],\ST[11])\downarrow f(\ST[6],\ST[8])\downarrow
f(\ST[10],\ST[4])\]
and
\[\ST[12]\]
\[=f(l_{(12,1)},r_{(12,1)}) \downarrow f(l_{(12,2)},r_{(12,2)})\]
\[=f(\ST[3],\ST[9]) \downarrow f(\ST[8],\ST[5]).\]
To solve such defined MCM problem by the pipeline implementation on
GPU, we have to modify the pipeline algorithm designed for the S-DP
problem, because the offset numbers for each \ST[$i$] may differ from
others.  Thus, for the MCM problem, we propose a modified pipeline
algorithm in Fig. \ref{fig:pipeline-alg-MCM}.  In the MCM-pipeline
algorithm, each element of \ST\ is computed by equation (\ref{eq:1}),
and \ST$[1]$, \ST[2],$\ldots$ ,\ST$[n]$ are preset with initial
values.  In substep 1, 2, and 3, the computation of
$f(l_{(i,j)},r_{(i,j)})$ is executed, and in substep 4, that obtained
value is used for the computation by $\downarrow$ and the result is
stored to \ST$[i]$.

\begin{figure}[h]
  {\hrulefill
    \ \\

    \underline{\bf A Pipeline Algorithm for MCM Problem}\\
  
    \ \ \ \FOR\ {\tt i} = $n+1$ \TO\ $n(n+1)/2 + n - 2$ \DO
    \begin{quote}
      \FOR\ {\tt j} = $1$ \TO\ $(n-1)$ \DO\ \underline{\bf in parallel}
      \begin{quote}
        Thread {\tt j} executes the following operation if $i_{\tt j} \leq k_{i_{\tt
            j}}$ where $i_{\tt j}={\tt i}-{\tt j}+1$:\\
         \noindent
         \hspace*{1em}{\it (substep 1)}
         \[{\tt v_l} = l_{(i_{\tt j},j)}; \]
         \hspace*{1em}{\it (substep 2)}
         \[{\tt v_r} = r_{(i_{\tt j},j)}; \]
         \hspace*{1em}{\it (substep 3)}
         \[{\tt v_S}= f({\tt v_l},{\tt v_r});  \]
         \hspace*{1em}{\it (substep 4)}
         \[
         \ST[i_{\rm j}] = \begin{cases}
           {\tt v_S}; \;\;({\rm j}=1) \\
           \ST[i_{\rm j}] \downarrow {\tt v_S}; \;\; ({\tt j}>1)
         \end{cases}
         \]
         where $\tt v_l, v_r,$ and $\tt v_S$ are local variables in
         a thread.
      \end{quote}
    \end{quote}
  \hrulefill
  }
  \caption{A pipeline algorithm for MCM problem}
  \label{fig:pipeline-alg-MCM}
\end{figure}

\subsection{Conflict-Free Memory Access of MCM Algorithm}

In this subsection, we prove that no memory access conflict occurs
during the execution of the MCM-pipeline algorithm described in Fig.
\ref{fig:pipeline-alg-MCM}.

To begin with, we prove the following lemma.
\begin{lemma}
  In substep 1 of an execution of the inner loop-body of the MCM
  algorithm, each thread has access to a distinct element of the array
  \ST.
\end{lemma} {\it Proof:} Assume that threads $p$ and $q$ try to read
the same element of \ST\ in substep 1 and that $p <q$ holds.  Let
$(row_p, col_p)$ (resp. $(row_q, col_q)$) be the pair of row and column
indexes of the elements in the original two-dimensional solution table
of triangle shape of the MCM problem for which thread $p$ (resp. $q$)
is now computing.  Since threads $p$ and $q$ try to read the same
element of \ST\ and in substep 1 they read the value for the left
argument of the function $f$, the relation $row_p =row_q$ must hold.
Then, since threads $p$ and $q$ respectively read the $p$-th and
$q$-th elements from the left in the same row of the original
two-dimensional solution table, the relation $p=q$ must hold if the
two threads read the same element of \ST, which contradicts the
assumption $p<q$.  \QEDPR

Next, we prove the following lemma, which can be proved in a similar
way to the proof of Lemma 1.
\begin{lemma}
  In substep 2 of an execution of the inner loop-body of the MCM
  algorithm, each thread has access to a distinct element of the array
  \ST.
\end{lemma} {\it Proof:} Assume that threads $p$ and $q$ try to read
the same element of \ST\ in substep 2 and that $p <q$ holds.  Let
$(row_p, col_p)$ (resp. $(row_q, col_q)$) be the pair of row and column
indexes of the elements in the original two-dimensional solution table
of triangle shape of the MCM problem for which thread $p$ (resp. $q$)
is now computing.  Since threads $p$ and $q$ try to read the same
element of \ST\ and in substep 2 they read the value for the right
argument of the function $f$, the relation $col_p =col_q$ must hold.
On the other hand, thread $p$ reads the $p$-th elements below of the
row $row_p$ of the original two-dimensional solution table, and thread
$q$ does the $q$-th elements below of the row $row_q$ of the table.
Since threads $p$ and $q$ read row $(row_p+p)$ and row $(row_q+q)$
respectively, the relation $row_p+p = row_q+q$ must hold if the two
threads read the same element of \ST.  Since $p<q$ and $col_p=col_q$
hold, the relation $row_p < row_q$ must hold from the way of mapping
from the original two-dimensional solution table to the linear array \ST.
This leads to the relation $row_p+p \not = row_q+q$, which contradicts
the assumption that threads $p$ and $q$ read the same element of array
\ST.
\QEDPR

In substep 3, each thread simply executes computation using local
variables.  In substep 4, it is obvious that each thread has access to
a distinct element of \ST.  Therefore, by Lemma 1 and Lemma 2, we
obtain the following theorem.
\begin{theorem}
  No memory access conflict occurs during the execution of the
  MCM-pipeline algorithm described in Fig. \ref{fig:pipeline-alg-MCM}.\QEDPR
\end{theorem}

As for the time-complexity of the MCM-pipeline implementation, from a
theoretical viewpoint, it takes only $O(n^2)$ steps with $(n-1)$
threads, because the outer loop takes $O(n^2)$ cycles and the inner
loop requires $O(1)$ time (Theorem 1).

\section{Concluding Remarks}
In this study, we examined the effectiveness of pipeline
implementations of Dynamic Programming (DP) on GPU. As an example, we
explained how to solve a matrix-chain multiplication (MCM) problem by
DP on GPU.  This problem can be sequentially solved in $O(n^3)$ steps
by DP where $n$ is the number of matrices.  In our approach, we solve
the MCM problem on GPU in a pipeline fashion, i.e., we use GPU cores
for supporting pipeline-stages so that many elements of the solution
table are partially computed in parallel at one time.  Since our
implementation determines one output value per one computational step
with $O(n)$ threads, we can solve the MCM problem in $O(n^2)$ steps
using $O(n)$ threads, which is an ideal speedup from the $O(n^3)$-step
sequential DP algorithm.

For future work, we plan to evaluate the performance of our pipeline
implementations.  From the experimental results shown in Section III,
it is obvious that the ideal speed up is not attained here.  This is
mainly due to the limitations on the bandwidth of memory on GPU.  That
is, as the problem size becomes large, all threads cannot always have
access to the target memory at one time, because unavoidable access
conflicts occur.  We also plan to study the relation between the
memory bandwidth and the performance of our pipeline implementation on
some theoretical GPU models (e.g., \cite{6799118}).

\newpage
\section*{Acknowledgment}
This paper is an extended version of our conference paper
\cite{miyazaki2018} where we proposed a pipeline-implementation for
the S-DP problem and discussed the memory access conflict issues.  In
this paper, we provide the MCM algorithm in details and formally prove
lemmas and theorem in Section IV.

\bibliographystyle{unsrt}

\end{document}